# Correlation function as a measure of the structure

O. Buryak,[1,2] A. Doroshkevich,[1,3]

1. Keldysh Inst. of Appl Math., Miusskaya pl.4, 125047 Moscow, Russia
2. University of Newcastle Upon Tyne, Newcastle, NE1 7RU, England
3. TAC, Blegdamsvej 17, DK-2100 Copenhagen, Ø Denmark

**Abstract.** Geometrical model of structure of the universe is examined to obtain analytical expression for the two points nonlinear correlation function. According to the model the objects (galaxies) are concentrated into two types of structure elements - filaments and sheets. We considered the filaments ( similar to galaxy filaments ) simply as straight lines and the sheets ( similar to superclusters of galaxies ) simply as planes. The homogeneously distributed objects are also taken into consideration. The spatial distribution of lines, planes and points is uncorrelated.

The nonlinear correlation function depends on four parameters and is similar to the observed and simulated ones for different samples. It describes quite well the correlation of galaxies, clusters of galaxies and dark matter distribution.

Possible interpretation of the parameters of nonlinear correlation function is discussed.

## 1. Introduction

The two point correlation function $\xi(|\mathbf{r}_1 - \mathbf{r}_2|) = \xi(r)$ is the most popular tool for the investigation of spatial object distribution both in the observational and simulated catalogues. The great potential of this method is known and it can be illustrated by the new interesting results obtained at the last years. Thus, Jorgensen et al. (1993, 1994) shown that oscillations in the spectra of primordial perturbations give rise to some interesting peculiarities in the correlation function. Let us remind also the paper of Mo et al., (1992) where characteristic scales were found in observational catalogues. Recently Peacock & Dodds (1994) reconstructed the initial power spectrum using the correlation analysis of different samples of galaxies and galaxy clusters. This list can be continued.

*Send offprint requests to*: A. Doroshkevich

The main drawback of correlation method is, of course, its phenomenological character. Indeed, the interpretation of nonlinear correlation function, $\xi \geq 1$, is ambiguous. The same correlation function can be generated by different spatial point distributions. For example, the similar correlation functions are generated by Soneira-Peebles (1978) model, by Voronoi tessellation (van de Weygaert, 1991), by the 'shell' model (Bahcall et al., 1989) and by many simulations with HDM, CDM and other power spectra. It is important that the spatial matter distribution in these models is very different. However the shape of nonlinear correlation functions is the same for a wide class of power spectra (Demianski & Doroshkevich (1992)).

The main properties of observed correlation function still have no physical explanation. Thus, for all observational samples the shape of the correlation function (for galaxies and clusters of galaxies) is approximately the same:

$$\xi(r) = (r/r_c)^{-\gamma} \qquad (1.1)$$

with the power index $\gamma \approx 1.8$. However, the correlation radius, $r_c$, as a rule, depends on the sample. Thus, for IRAS, CfA and APM catalogues $r_c \approx (5 - 6)\ h^{-1}Mpc$ (Davis et al., 1988, Loveday et al., 1992). But for the sample of dwarf galaxies (Thuan et al., 1991) $r_c \approx 4\ h^{-1}Mpc$ and $r_c \approx 15 h^{-1}Mpc$ for the Great Wall region (Ramella, Heller & Huchra, 1992). Of course, these variations are produced by strong peculiarities of the samples. However, adequate physical explanation of these phenomena are still lacking.

Let us also note that the strong correlation of clusters of galaxies has no physical explanation. Of course, similar correlation has been simulated (see, e.g., Klypin & Rhee, 1994, Croft & Efstathiou, 1994) and, therefore, the parameters of correlation function can be caused by the power spectrum. However, at present there is a strong limit to what can be done in this area.

There are different interpretations of the correlation function. Thus, it can be considered as the evidence in favor of hierarchical clustering (Soneira & Peebles, 1978) or in favor of fractal character of galaxy distribution (Coleman & Pietronero, 1992). Here we interpret the shape of correlation function as the evidence in favor of strong concentration of galaxies in structure elements. Evidently, additional information is necessary for a proper interpretation of the correlation function. Nevertheless, the correlation analysis gives the important quantitative information about the spatial matter distribution and it cannot be ignored in theoretical considerations.

The simplest model of the nonlinear correlation function is provided by a geometrical model. According to this model the nonlinear correlation function is generated by the concentration of significant fraction of the objects along lines and planes. The great potential of this approach was emphasized by Zeldovich (see, e.g. Shandarin et al. 1983). This model is suitable to describe the Large and Super Large Scale Structure (LSS and SLSS) in the observed distribution of galaxies in deep galaxy surveys and the dark matter (DM) distribution in dynamical N-body simulations (Buryak, Doroshkevich & Fong, 1994, hereafter BDF, Doroshkevich et al., 1995, hereafter DFGMM, Doroshkevich, 1995). In the following we use this model to construct the nonlinear correlation function.

In the geometrical model filaments are considered simply as lines and sheets simply as planes. To be more specific, we assume that a filament is identified with a line segment and a sheet corresponds to a finite planar region. If filaments and sheets are large enough we neglect the edge effects in our analysis.

This model can be employed to describe the spatial distribution of various objects. Thus, for the observed catalogues a 'filament' is represented by a chain of galaxies or clusters of galaxy, a 'sheet' is represented by superclusters or richest filaments of galaxies. For the DM distribution in simulated catalogues a 'filament' can be identified with strongly extended over dense region, whereas a 'sheet' with high dense wall at the boundary of voids or under dense regions.

This approach is not general and it has its limitations. It is an intermediate step between the local description of matter distribution by means of density and velocity fields and such global methods as the percolation or fractal analyses. The geometrical model works at scales comparable to the correlation radius and, therefore, it can be used to describe the nonlinear correlation function.

Random distribution of straight lines can be characterized by the surface density, $\sigma_f$, i.e., the mean number of lines intersecting a unit area of arbitrary orientation. Similarly, random distribution of sheets can be characterized by the linear density, $\sigma_s$, i.e., the mean number density of sheets crossing an arbitrary straight line. Of course, we have to include also some fraction of homogeneously distributed points. It is characterized by the 3D number density, $n_h$. Parameters of the correlation function - the power index, $\gamma$, the correlation radius, $r_c$, and the zero point, $r_0$, - are related to structure parameters.

In the geometrical model the three and four point correlation functions can be analytically calculated as well. However, these expressions are not so simple and, actually, the main advantage of analytical approach is lost. Simple numerical simulations of such matter distribution seem to be more perspective ways to examine the three and four point correlation functions.

This paper is organized as follows. In section 2 the model of the correlation function is formulated. In Section 3 we examine this model and obtain the two point correlation function of galaxy distribution. In Section 4 we reconstruct the DM correlation function using some information about the DM distribution from dynamical N-body simulations. In Section 5 we propose a set of models for the cluster-cluster correlation function. In Section 6 and 7 the evolution of correlation function and our results are discussed.

## 2. The analytical model of correlation function

In this section we consider the correlation function for three simple objects distributions: a homogeneous point distribution and randomly distributed straight lines and planes. Further we construct the composite correlation function including all three types of point distribution.

### 2.1. Correlation function of homogeneous point distribution

The simplest model of the object distribution is a homogeneous Poissonian distribution with a mean number density $<n_g> = n_h$. As it is known, in this case the number of actual pairs, DD, at the distance $(r, r+dr)$ is identical to the number of Poissonian pairs, PP, at the same distance

$$DD = PP = <n_g> 4\pi r^2 dr = 4\pi n_h r^2 dr \quad (2.1)$$

Therefore, in this model we obtain the trivial result:

$$1 + \xi(r) = \frac{DD}{PP} = 1$$
$$\xi(r) = 0 \quad (2.2)$$

### 2.2. Correlation function of randomly distributed straight lines

The more interesting model is a system of randomly distributed identical straight lines. It is characterized by the surface density of lines, $\sigma_f$. We assume that objects are distributed homogeneously along lines with a linear number density $n_1$.

Let us consider a sphere with a radius $r$ which has an object and, therefore, a straight line at its center. In thus case the main characteristics of the point distribution are: the mean number of lines, $N_f$, intersecting the sphere, the mean length of intersection, $l_f$, and the mean 3D number density of objects, $<n_g>$ inside the sphere. These parameters can be found by the standard method (see, e.g., BDF). Here we present only the final expressions:

$$N_f = 0.5\sigma_f S = 2\pi r^2 \sigma_f, \quad l_f = 4V/S = \frac{4}{3}r$$
$$<n> = n_1 l_f N_f / V = 2\sigma_f n_1 \quad (2.3)$$

where $V$ and $S$ are the volume and surface area of the sphere.

However, it must be taken into account that a line is crossing through the center of sphere. Thus, we have to use the conditional mean number of lines intersecting the sphere. This value, $N_f^*$, can be written in the following form:

$$N_f^* = \frac{1}{(1-W_0)} \sum_{k=0}^{\infty} k W_{k+1}(N_f) = \frac{N_f}{(1-W_0)} - 1 \quad (2.4)$$

where $W_k(N_f)$ is a probability to find $k$ lines intersecting the sphere. Therefore, using the Poissonian probability $W_k = \frac{N_f^k}{k!} e^{-N_f}$ we obtain:

$$DD = 2n_1 dr + \frac{4}{3} n_1 dr N_f^*,$$

$$PP = 8\pi r^2 \sigma_f n_1 dr = 4 N_f n_1 dr,$$

$$\xi(r) = \frac{1}{6N_f} \left(1 + \frac{2N_f}{e^{N_f} - 1}\right) - \frac{2}{3}. \quad (2.5)$$

Thus, for small $r$ ($N_f \ll 1$) $\xi = (2N_f)^{-1} \propto r^{-2}$, and for large $r$ ($N_f \gg 1$) $\xi(r) \approx -2/3$.

*2.3. Correlation function of randomly distributed sheets.*

This case can be treated in the same way as a previous one. We assume that the objects are distributed homogeneously on identical sheets with a surface density $n_2$.

Let us consider again a sphere with radius $r$ which has in its center an object and, therefore, a sheet. For this model the mean number of sheets, $N_s$, intersecting the sphere, the mean area of intersection, $S_s$, and the mean 3D number density of objects inside the sphere, $<n_g>$, are:

$$N_s = 4\sigma_s r, \quad S_s = \frac{2}{3}\pi r^2, \quad <n_g> = 2\sigma_s n_2 \quad (2.6)$$

The conditional mean number of sheets intersecting the sphere, $N_s^*$, is

$$N_s^* = \frac{1}{(1-W_0)} \sum_{k=0}^{\infty} k W_{k+1}(N_s) = \frac{N_s}{(1-W_0)} - 1 \quad (2.7)$$

Therefore, in this model

$$DD = 2\pi r dr n_2 + \frac{4}{3} \pi r dr n_2 N_s^*,$$

$$PP = 8\pi r^2 dr \sigma_s n_2 = 4 N_s n_2 \pi r dr,$$

$$\xi(r) = \frac{1}{3N_s} \left(1 + \frac{2N_s}{e^{N_s} - 1}\right) - \frac{1}{3}, \quad (2.8)$$

and, therefore, $\xi = 1/N_s \propto r^{-1}$ for small $r$ ($N_s \ll 1$), and $\xi(r) \approx -1/3$ for large $r$ ($N_s \gg 1$)

*2.4. The composite model of correlation function.*

It is clear that the simple models considered above cannot be used to describe any realistic catalogues. It is more reasonable to use composite model in which all three types of structure elements (sheet, filament and homogeneous populations) are mixed with the fraction factors $f_s$, $f_f$ and $f_h$ correspondingly.

The mean density of particles and fraction factors can be written as

$$<n> = n_h + 2\sigma_f n_1 + 2\sigma_s n_2 = n_h + n_f + n_s \quad (2.9)$$

$$f_h = n_h/<n>, \quad f_f = 2\sigma_f n_1/<n>, \quad f_s = 2\sigma_s n_2/<n>$$

Therefore, for the composite model we obtain

$$DD = f_h \left(4\pi r^2 dr n_h + \frac{4}{3} n_1 dr N_f + \frac{4}{3}\pi r dr n_2 N_s\right)$$

$$+ f_f \left(2 n_1 dr + \frac{4}{3} n_1 dr N_f^* + \frac{4}{3}\pi r dr n_2 N_s + 4\pi r^2 dr n_h\right)$$

$$+ f_s \left(2\pi r dr n_2 + \frac{4}{3} n_1 dr N_f + \frac{4}{3}\pi r dr n_2 N_s^* + 4\pi r^2 dr n_h\right)$$

$$PP = 4\pi r^2 dr <n> = 4\pi r^2 dr n_h / f_h$$

$$= 4\pi r^2 dr n_f / f_f = 4\pi r^2 dr n_s / f_s$$

where it is assumed that a) the sphere is centered on an object from the Poisson population, and b) the sphere is centered on an object from filament and sheet populations. Finally, the analytical expression for the correlation function is

$$\xi(r) = \frac{f_f^2}{6N_f}\left(1 + \frac{2N_f}{e^{N_f}-1}\right) + \frac{f_s^2}{3N_s}\left(1 + \frac{2N_s}{e^{N_s}-1}\right)$$

$$- (f_s + 2f_f)/3 \quad (2.10)$$

$$f_h + f_f + f_s = 1$$

Let us remind that according to (2.3) and (2.6) $N_f = 2\pi r^2 \sigma_f$, $N_s = 4\sigma_s r$ and, therefore, for small $r$ $\xi(r) \propto r^{-2}$. This composite model reduces to (2.2), (2.5) and (2.8) when $f_h = 1$, $f_f = 1$ and $f_s = 1$ correspondingly.

*2.5. Some properties of the correlation function.*

Let us consider in more detail the relationship (2.10) and find the main properties of correlation function, $\xi(r)$. Naturally, our model cannot be applied for large distances, $r \gg r_c$, $\xi(r) \ll 1$, where the matter distribution 'remembers' the initial perturbations and, therefore, the small correlations in the distribution of structure elements are essential. Let us remind that we assumed the uncorrelated spatial distribution of structure elements and, so, the correlation function (2.10) describes only the matter concentration into structure elements.

Thus, we will concentrate our attention on the nonlinear region where the correlation function is large enough and our approach is justified. As a rule, in this region, for the analytical consideration we can use the simpler version of (2.10):

$$\xi(r) = (r_f/r)^2 + 2(r_s/r) - \xi_\infty \quad (2.11)$$

$$r_f^2 = f_f^2/4\pi\sigma_f, \quad r_s = f_s^2/8\sigma_s, \quad \xi_\infty = (f_s + 2f_f)/3$$

where we assumed that $N_f \leq 1$, $N_s \leq 1$.

The correlation radius, $r_c$, is defined by the relation

$$\xi(r_c) = 1, \quad r_c = \frac{r_s}{(1+\xi_\infty)}\left(1 + \sqrt{1+\zeta^2}\right), \quad (2.12)$$

$$\zeta^2 = (1+\xi_\infty)(r_f/r_s)^2.$$

Another important parameter of the correlation function, the zero point, $r_0$, can be found from

$$\xi(r_0) = 0, \quad r_0 = \frac{r_s}{\xi_\infty}\left(1 + \sqrt{1 + \frac{\xi_\infty}{1+\xi_\infty}\zeta^2}\right) \quad (2.13)$$

For the important case $r_s \gg r_f$, $\zeta \ll 1$ these expressions simplify to

$$r_c \approx 2r_s/(1+\xi_\infty), \quad r_0 \approx 2r_s/\xi_\infty$$

$$r_0/r_c - 1 \approx 1/\xi_\infty \quad (2.14)$$

It establishes the interesting relations between the main parameters of the correlation function and characteristics of the structure.

For the model (2.11) near the point $r = r_c$ the power index of the correlation function $\gamma$ is a weak function of $r$:

$$\gamma = \gamma_c(1 + \beta_c(\frac{r_c}{r} - 1) + ...) \quad (2.16)$$

$$\gamma_c = 2(1+\xi_\infty)\frac{\sqrt{1+\zeta^2}}{\sqrt{1+\zeta^2}+1}, \quad 1+\xi_\infty \le \gamma_c \le 2(1+\xi_\infty)$$

$$\beta_c = \sqrt{\gamma_c - 1 - \xi_\infty}\,(3 - \gamma_c - 2(1+\xi_\infty)/\gamma_c)$$

It is interesting that in two limiting cases, $\zeta \ll 1$ and $\zeta \gg 1$ the power index is connected with $r_c$ and $r_0$ as follows:

$$\gamma_c \approx (1 - r_c/r_0)^{-1}, \quad \zeta \ll 1 \quad (2.17a)$$

$$\gamma_c \approx 2(1 - r_c^2/r_0^2)^{-1}, \quad \zeta \gg 1 \quad (2.17b)$$

## 3. The correlation function of galaxy distribution

In this section we consider the composite model (2.10) with the structure parameters found from the observed galaxy distribution. We use the estimates of the main structure parameters obtained by BDF from deep galaxy surveys.

Thus, according to BDF, the distribution of galaxy filaments can be characterized by the surface density

$$\sigma_f \approx 0.5 \times 10^{-2} h^2 Mpc^{-2} \quad (3.1)$$

In this case the SLSS elements can be considered as the sheet population. The characteristic scale of the SLSS observed from two different types of the surveys gives:

$$\sigma_s \approx 2 \times 10^{-2} h Mpc^{-1}, \quad \sigma_s \approx 0.77 \times 10^{-2} h Mpc^{-1}. \quad (3.2)$$

To test the possible shape of the composite correlation function for these structure parameters two models with the same value of $\sigma_f$ and different parameters $\sigma_s$, $f_f$ and $f_s$ were calculated. For both models we assumed that all galaxies are concentrated into the structure elements and, thus, $f_h = 0$.

The functions $\xi(r)$ are shown in the Fig. 1 for the correlation radius $r_c = 5.5 h^{-1} Mpc$. The parameters of the models are presented in Table I. The straight line in Fig. 1 corresponds to the function

$$\xi_{gal}(r) = (r_c/r)^{1.8} \quad (3.3)$$

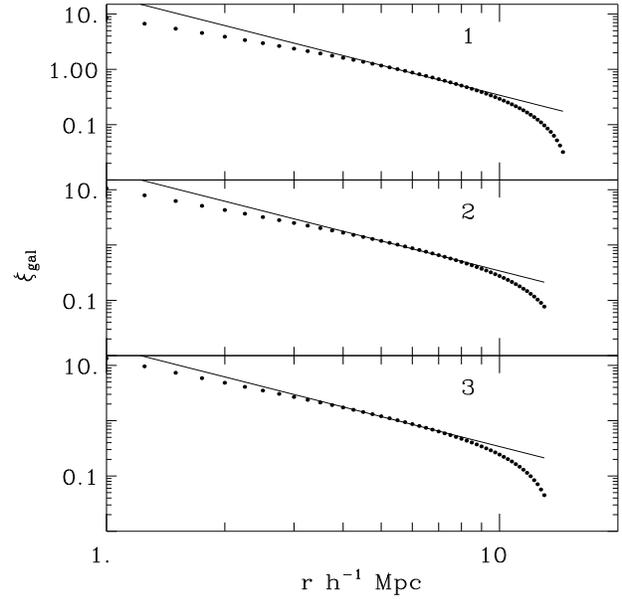

**Fig. 1.** Correlation function (2.10) for galaxies. Parameters are presented in Table I. Straight lines correspond to the function (3.3).

These results illustrate the potential of the composite model and allow us to evaluate the galaxy fractions $f_s$ and $f_f$ in different elements of the structure.

Table I

| $\sigma_s 10^{-2}$ $hMpc^{-1}$ | $f_s$ | $n_2 h$ $Mpc^{-2}$ | $\sigma_f 10^{-2}$ $h^2 Mpc^{-2}$ | $f_f$ | $n_1 h$ $Mpc^{-1}$ | $r_0 h^{-1}$ $Mpc$ |
|---|---|---|---|---|---|---|
| 2. | 0.84 | 0.21 | 0.50 | 0.16 | 0.16 | 15.2 |
| 1. | 0.58 | 0.29 | 0.50 | 0.42 | 0.42 | 14.8 |
| 1. | 0.56 | 0.28 | 0.25 | 0.44 | 0.88 | 13.8 |

There is however a disagreement between estimates (3.1), (3.2) and assumptions of the model (2.10). The value of the filament surface density, $\sigma_f$, used above was obtained by BDF as a low limit of size of the structure when the poorest filaments were taken into account. However, in the model (2.10) all filaments are considered as identical and, therefore, the surface density of the 'average' filaments should be used. It could be smaller by a factor of 1.5 - 2 than the accepted value. To test such 'correction' in Fig. 1 the correlation function for the smaller surface density $\sigma_f$ ( the model 3 in Table I ) is shown as well. In this case the correlation radius $r_c = 5.5 h^{-1} Mpc$ but the fractions $f_s$ and $f_f$ have different values than above.

Other values of the correlation radius $r_c$ can be obtained for other values of the fractions $f_f$, $f_s$. We conclude that, actually, the correlation radius depends on many structure parameters in the investigated catalogue.

The estimates of the mean linear density of galaxies in filaments, $n_1$, and surface density of galaxies in sheets, $n_2$ are also presented in Table I for $<n_g> \approx 10^{-2} h^3 Mpc^{-3}$. The mean separation of galaxies in the structure elements, $n_1^{-1}$ and $n_2^{-1/2}$ are less than $<n>^{-1/3}$ by a factor of 2-5.

## 4. The correlation function of dark matter distribution.

For simulated DM distribution, as a rule, the correlation function of DM spatial distribution is characterized by the same power index, $\gamma \approx 1.8$, as for galaxies and the correlation radius, $r_c \approx 3.h^{-1} Mpc$. Usually this difference in correlation radius is attributed to the 'biasing' factor, $b$. Analysis of LSS and SLSS in numerical simulations (DFGMM) shows however that the spatial distribution of 'galaxies' and DM is very different. Therefore, it is interesting to test consistency between the parameters of the correlation function and DM structure on the basis of model (2.10).

As it was found by DFGMM the DM distribution is dominated by the sheet like structure. Filamentary structure exists mainly in the smallest scale and it includes only a small fraction of the DM population. But, unlike in the galaxy distribution, the fraction of homogeneously distributed DM particles could be significant (10-20%).

Thus, properties and parameters of the LSS for 'galaxy' and DM populations are very different. However, the shape and the power index of the correlation function are the same for both populations. The model (2.10) allows to obtain the correlation function with the reasonable correlation radius, $r_c \approx 3.h^{-1} Mpc$, and with the power index, $\gamma \approx 1.8$ for the range parameters presented in Table II.

Table II

| $\sigma_s 10^{-2}$ $hMpc^{-1}$ | $f_s$ | $\sigma_f 10^{-2}$ $h^2 Mpc^{-2}$ | $f_f$ | $r_c h^{-1}$ $Mpc$ | $r_0 h^{-1}$ $Mpc$ |
|---|---|---|---|---|---|
| 4. | 0.88 | 1.0 | 0.12 | 3.0 | 8.5 |
| 4. | 0.39 | 0.25 | 0.61 | 3.0 | 5.2 |
| 3. | 0.72 | 0.25 | 0.03 | 3.0 | 11.0 |

Estimates of $\sigma_s$ and $\sigma_f$ found by DFGMM were used. The corresponding correlation functions are shown in Fig. 2.

The straight line in Fig. 2 corresponds to the function

$$\xi_{DM}(r) = (r_c/r)^{1.8} \qquad (4.1)$$

with $r_c = 3.h^{-1} Mpc$.

Notice again that in the model (2.10) all filaments and sheets were assumed to be identical while the direct analysis (DFGMM) shows distribution of the structure elements with richness. Thus, the model (2.10) should be also corrected for this factor. However, in this paper we are interested in dependence of the correlation function on other structure parameters. For example, the model 2 in Fig. 2 and in Table II corresponds to the smaller surface density of DM filaments. The model 3 in Fig. 2 and in Table II illustrates the influence of the homogeneous background of DM particles ($f_h = 0.25$) on the correlation function.

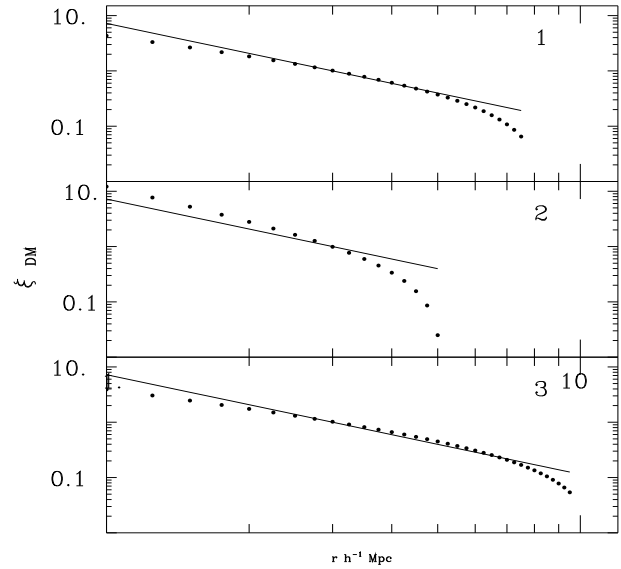

**Fig. 2.** Correlation function (2.10) for DM distribution. Parameters are presented in Table II. Straight lines correspond to the function (4.1).

This sample of correlation functions demonstrates that the same correlation function can be generated by different DM distribution.

## 5. The correlation function of clusters of galaxies.

The observed cluster-cluster correlation function is characterized by the power index, $\gamma$, by the correlation radius, $r_c$, and by the zero point, $r_0$:

$$\gamma \approx 1.8, \quad r_c \approx 20 h^{-1} Mpc, \quad r_0 \approx (50-60) h^{-1} Mpc \qquad (5.1)$$

(see, e.g., Dalton et al., 1992, Peacock & West, 1992, Scaramella et al., 1993). The mean number density of clusters, $n_{cl}$, and the mean separation of the clusters, $D_{cl}$, are

$$n_{cl} \approx 2.4 \times 10^{-5} h^3 Mpc^{-3},$$

$$D_{cl} = n_{cl}^{-1/3} \approx 35 h^{-1} Mpc. \qquad (5.2)$$

The values of $r_c$, $r_0$, $n_{cl}$ and $D_{cl}$ strongly depend on the sample. There are also other estimates (Bahcall, 1988, Postman, Huchra & Geller, 1992)

$$n_{cl} \approx 6 \times 10^{-6} h^3 Mpc^{-3}, \quad D_{cl} \approx 55 h^{-1} Mpc. \qquad (5.3)$$

Estimates of the correlation radius, $r_c$, and the zero point, $r_0$, for some samples may be as high as $50 h^{-1} Mpc$ and $100 h^{-1} Mpc$ correspondingly (Bahcall, 1988).

Table III

| $\sigma_s 10^{-2}$ $hMpc^{-1}$ | $f_s$ | $n_2 10^{-2}$ $h^2 Mpc^{-2}$ | $\sigma_f 10^{-4}$ $h^2 Mpc^{-2}$ | $f_f$ | $n_1 h$ $Mpc^{-1}$ | $r_0 h^{-1}$ $Mpc$ |
|---|---|---|---|---|---|---|
| 0.20 | 0.39 | 0.24 | 0.64 | 0.41 | 0.08 | 51. |
| 0.20 | 0.26 | 0.16 | 0.25 | 0.34 | 0.16 | 46. |
| 0.20 | 0.36 | 0.21 | 0.16 | 0.20 | 0.15 | 64. |
| 0.15 | 0.27 | 0.22 | 0.09 | 0.17 | 0.23 | 68. |
| 0.10 | 0.27 | 0.32 | 0.50 | 0.33 | 0.08 | 60. |

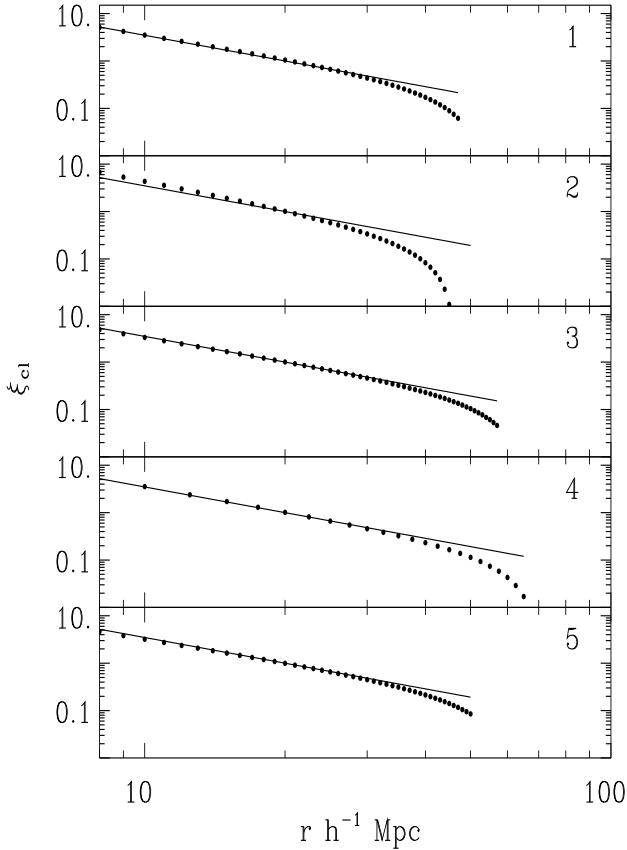

**Fig. 3.** Correlation function for the clusters of galaxies. Parameters are presented in Table III. Straight lines correspond to the function $\xi_{cl} = (r_c/r)^{1.8}$ with $r_c = 20h^{-1}Mpc$.

Significant fraction of clusters forms chains in superclusters of galaxies which are similar to the well known chain in the Perseus supercluster (rich clusters Abel 426, 347 & 262). An example of clusters walls has been demonstrated by Kopylov et al., (1988). A catalog of clusters of clusters of galaxies was prepared by Bahcall & Soneira (1984). The linear size of such clusters of clusters could be as large as $100h^{-1}Mpc$.

On the other hand, the percolation analysis gives the value $P \approx 3$ (Klypin, 1988) for the percolation parameter which is close to the Poissonian one. This fact shows that the cluster population has a less pronounced structure and there is a significant fraction of isolated clusters.

Therefore, it can be expected that the spatial distribution of clusters can be described by our analytical model. It is significant that a numerical 'shell' model (Bahcall et al, 1989) which is close to the model (2.10) in many respects is consistent with the observed cluster correlation. Moreover, the observed parameters $r_c$, $r_0$ and $\gamma$ (5.1) approximately satisfy the relationship (2.17a).

We present here five models of correlation function. Their parameters are listed in Table III. Fig. 3 shows the corresponding correlation functions. For all models the correlation radius is equal to $20h^{-1}Mpc$, and we used data (5.2) to obtain the mean number density of clusters in the structure elements. These models show that, indeed, it is possible to interpret the observed and simulated cluster-cluster correlation functions as a result of the cluster concentrations into structure elements of the SLSS.

Of course, the spatial distribution of clusters should be investigated in more detail to reveal the quantitative characteristics of their distribution. However, the small values of $\sigma_f$ and $\sigma_s$ in Table III show that it is necessary to use very large catalogues in order to get statistically significant estimates. On the other hand, found here mean separation of clusters in sheets $D_{cl}^s \approx (20 - 25)h^{-1}Mpc$ is close to that in the homogeneous population

$$D_{cl}^h \approx 35 f_h^{-1/3} h^{-1} Mpc \approx 45 h^{-1} Mpc. \qquad (5.4)$$

This means that it is very difficult to identify sheets in observational catalogues. For the filament population the expected mean separation of clusters is close to $5h^{-1}Mpc$ and it is more easier to identify it.

## 6. Evolution of the correlation function.

Evolution of the correlation function has been simulated for different spectra of the primordial perturbations and, in general, it can be described as a growth of amplitude (or correlation radius, $r_c$) and power index, $\gamma$. On the other hand, the direct analysis of the DM structure evolution (DFGMM) shows that there are three main evolutionary periods:

1. The structure arises during a short period which is defined by the amplitude of density perturbations. At the same time the nonlinear part of correlation function is formed.

2. Later the typical scale of the LSS grows slowly.

3. During the last period of evolution sheets are transformed to filaments and further to a system of knots.

Therefore, as a rule, $f_f$ is an increasing functions of time while $f_s$ is decreasing one. Expressions (2.12), (2.16) show that this picture of structure evolution is in a qualitative agreement with evolution of the correlation function. However, the presently available data does not allow us to test this evolutionary history more quantitatively.

For the cluster - cluster correlation function the evolution is very slow because the mean velocity of clusters is relatively small. Probably, the continuous formation of new clusters can be considered as an essential factor in the evolution of the spatial distribution of clusters.

## 7. Discussion.

The main aim of this paper is to show that the nonlinear part of observed and simulated correlation functions can be generated by the concentration of the objects (galaxies, DM particles and clusters of galaxies) into structure elements - filaments and sheets. Of course, the converse is not true and various spatial distributions of objects can generate the same correlation function. It is essential that in order to construct the model of observed correlation function we have to use no less than two types of structure elements (filaments and sheets) and, therefore, no less than three structure parameters. Probably, the nonhomogeneous matter distribution into structure elements produces similar effects in numerical models (Bahcall et al., 1989, van de Weygaert, 1991).

Thus, we can conclude that the parameters of the nonlinear correlation function depend on many structure parameters and, therefore, the interpretation of the correlation function is ambiguous.

The model (2.10) gives a good approximation of the observed correlation function in nonlinear area for a wide range of structure parameters. Moreover, using the observed values of basic structure parameters, such as the surface density of filaments, $\sigma_f$, and linear density of sheets, $\sigma_s$, we can also evaluate the fraction of objects concentrated in different types of the structure elements and find the reasonable mean separation of objects in structure elements. These results show that in many instances the model (2.10) corresponds to the actual situation.

The analysis of Sec. 3 & 4 shows that the biasing parameter, $b$, describing the difference between the spatial distribution of DM and 'galaxies' actually depends on many structure parameters. Therefore, the biasing parameter is not an universal characteristic of matter distribution (or DM model), but, rather, it is a characteristic of the formation of galaxies and structure. Thus, for galaxies and clusters the structure parameters have different origin and there is no reasons to expect the same biasing parameter for these objects. This conclusion is consistent with results of Peacock & Dodds (1994). Moreover, there are good reasons to think that the biasing parameter depends on the considered sample of objects and, so, it is different for over and under dense regions.

The model (2.10) is very simple and there is a limit to what can be done in its framework. Thus, this model cannot explain the remarkably stable value of the power index $\gamma = 1.8$ in all observed samples for all classes of objects. Of course, the DM and galaxy distributions evolve and therefore the value of $\gamma$ characterizes the evolutionary period. However, evolution of the cluster distribution is slow (Croft & Efstathiou, 1994) and, perhaps, it manifests in the continuous cluster formation. Thus, in this case the power index characterizes rather the spatial distribution of regions where clusters form and emergence of the same value of $\gamma$ is surprising.

As noted in the introduction, for some observed samples of galaxies the correlation radii are different although the power index is the same. On the basis of the model (2.10) this fact can be accounted for by changing the structure parameters while maintaining the same (or close) fractions $f_s$ and $f_f$. Probably, the universal dimensionless correlation function (Bahcall, 1988) can be interpreted by the same way. However, such explanation is not fully satisfactory and warrants further investigation.

Sometimes the power law dependence of the correlation function (1.1) and the surprisingly universal value of $\gamma \approx 1.8$ are considered as the evidence in favor of fractal and multifractal properties of the matter distribution (see, e.g., Coleman & Pietronero, 1992). Actually, such approach may be effectively used to obtain additional quantitative characteristics of the object distribution. However, the previous discussion shows that these facts are not a good basis for the modification of fundamental concepts of cosmology. Various evolutionary histories can result in similar correlation and fractal properties and therefore interpretation of these properties is ambiguous.

Traditionally, the strong correlation of clusters is described by the theory of linear Gaussian fluctuations in the smoothed density field (see, e.g., Holtzman & Primak, 1992, Croft & Efstathiou, 1994). However, such approach does not provide physical explanation of the strong cluster-cluster correlation and, so, it has rather phenomenological character.

However, the results of Sec. 5 show that the observed correlation function can be generated by inclusion of the clusters into structure elements mean separation of which exceeds the correlation radius by a factor of 5 - 10. In this way we have to consider range of scales which are typical for the spatial distribution of gravitational potential of perturbations and, thus, the strong correlation can be attributed to the concentration of clusters into elements of the SLSS.

Concentration of clusters in the structure elements is consistent with general ideas of the nonlinear theory of gravitational instability. Indeed, as it has been argued by Buryak et al. (1992) and Kofman et al. (1992), the spatial distribution of SLSS elements is determined by the spatial distribution of the gravitational (or velocity) potential. It is therefore reasonable to expect that the spatial distribution of rich clusters is also determined by the gravitational potential of primordial perturbations. Rich clusters should be concentrated mainly into a system of 2D sheet and 1D filament inside deep 'wells' or, partly, around very deep minima of potential. If this assumption is correct then the SLSS and the structure in the spatial distribution of clusters are generated by the same causes. In this case the strong cluster - cluster correlation is simply a consequence of preferential cluster formation in the SLSS elements. To test this hypothesis we have to study the cross correlation of the potential and clusters distribution in numerical simulations. Such an analysis can also determine the fraction of clusters distributed quasi homogeneously.

However, although the concentration of clusters and galaxies into structure elements is similar, what was emphasized, for example, by the universal dimensionless correlation function (Bahcall, 1988 ), this analogy has rather formal character because the physical causes of the LSS and SLSS formation and their evolution are different.

*Acknowledgements* This paper was supported in part by Danmarks Grundforskningsfond through its support for an establishment of Theoretical Astrophysics Center. This work was also supported by grant INTAS-93-68 and by the Center of Cosmo-Particle Physics, Moscow.